# A Broadband Superabsorber at Optical Frequencies: Design and Demonstration

*Arvind Nagarajan [1,2,\*], Kumar Vivek [1,2,+], Manav Shah [3], Venu Gopal Achanta [3], and Giampiero Gerini [1,2]*

1) Electromagnetics Group, Technische Universiteit Eindhoven (TU/e), 5600 MB Eindhoven, The Netherlands.
2) Optics Department, Netherlands Organization for Applied Scientific Research (TNO), Stieltjesweg 1, 2628 CK Delft, The Netherlands.
3) FOTON Laboratory, Tata Institute of Fundamental Research, Homi Bhabha Road, Mumbai 400005, India.

[*] E-mail: arvind.nagarajan@tno.nl

[+] Currently at ASML, The Netherlands

**Abstract**: Metasurface based superabsorbers exhibit near unity absorbance. While the absorption peak can be tuned by the geometry/size of the sub-wavelength resonator, broadband absorption can be obtained by placing multiple resonators of various size/shapes in a unit cell. Metal dispersion hinders high performance broadband absorption at optical frequencies and careful designing is essential to achieve good structures. We propose a novel analytical framework for designing a broadband superabsorber which is much faster than the time consuming full wave simulations that are employed so far. Analytical expressions are derived for the wavelength dependency of the design parameters, which are then used in the optimization of broadband absorption. Numerical simulations report an average polarization-independent absorption of ~97% in the 450 – 950 nm spectral region with a near unity absorption (99.36%) in the 500 – 850 nm region. Experimentally, we demonstrate an average absorption over 98% in the 450-950 nm spectral region at 20° incident angle The designed superabsorber is polarization insensitive and has a weak launch angle dependency. The proposed framework simplifies the design process and provides a quicker optimal solution for high performance broadband superabsorbers

# 1. Introduction

Perfect absorption and wave trapping have been major challenges in photonics with numerous potential applications in energy harvesting [1] and high-performance sensing [2]. A promising pathway to achieve this is to use metasurfaces presenting additional degrees of freedom for tailoring the electromagnetic properties of materials [3]. Metallic metasurfaces, based on surface plasmon polariton interactions exhibit unique optical properties such as high transmission [4], high field concentration [5] and perfect absorption [6]. In the past decade, sub-wavelength patterned metal-dielectric-metal (MDM) superabsorbers (also termed as perfect absorbers) have been demonstrated to completely absorb the incident light by realizing a perfect impedance matching with free space and by exploiting very high loss guided transmission [7–9]. Superabsorbers can be classified into narrow-band absorbers [10] and broadband absorbers [11–13] depending on the absorption bandwidth. While narrow-band absorbers are generally used in sensing [14–16] and absorption filtering [17], broadband absorbers find application in thermo-photovoltaics [18,19] and anti-reflection surfaces [20–23]. Broadband absorption is realized by placing multiple resonators in a unit cell and have been demonstrated at various frequencies extending from optical to terahertz range [24–33]. Specifically, at optical frequencies, Azad et. al. [13] experimentally demonstrated a broadband, polarization independent, wide-angle MDM superabsorber accomplishing more than 90% absorption in the 350 – 1100 nm spectral region. Plasmonic nanocomposites [34,35], Carbon nanotubes [36], nanocone arrays [37,38], and lossy metals [39] are other alternative approaches to achieve broadband absorption at optical frequencies. However, these techniques either do not offer tunability of absorption bandwidth or have complicated fabrication procedures. On the other hand, nano patterned MDM surfaces offers the possibility to tailor the design for a specific resonance bandwidth and can be easily fabricated with existing nanolithography techniques. They can also be realized over large areas using nano-imprint lithography [40]. However, at optical frequencies, the metal dispersion hinders the optimization of resonator dimensions and geometrical features for broadband absorption. Almost all of the previous works on broadband MDM superabsorbers [11–13,24–33] rely on time consuming full-wave simulations requiring complex optimization routines such as Genetic algorithms [41], and reported broadband performances at optical frequencies are fewer compared to those at Microware or THz. There is a need for an analytical design methodology to get more insights into the wavelength dependency of the resonator dimensions and other key design parameters, and to allow a better optimization for broadband performance. At the moment, such an analytical methodology has not been reported.

In this paper, we present a novel analytical framework to design and optimize a broadband superabsorber at optical frequencies. The developed analytical framework, uses Gap-plasmon dispersion to get the dimensions of the resonators, and then treats the patterned MDM superabsorber as stacked layers, by homogenizing the array of resonators into an effective medium. The stacked layers are then solved using Transfer Matrix Method. Numerical simulations based on Finite Element Method (FEM) technique test the effectiveness of the proposed framework. The proposed MDM superabsorber is fabricated using electron beam lithography.

**2. Analytical framework**

A typical MDM superabsorber consists of an array of metal resonators separated from a ground metal plane by a thin dielectric layer. The array of resonators very efficiently couple the incoming light into surface plasmon polaritons (SPPs) at the metal-dielectric interface. The SPPs are then completely dissipated in the structure due to losses, leading to perfect absorption. The resonance can be tuned by modifying the size of the resonators, and broadband performance can be achieved by multiplexing multiple resonators of different shapes in the unit cell. A schematic of a MDM superabsorber is shown in the inset of **Figure 1**. Here W is the width of the resonator, H is the height and $t_d$ is the thickness of the dielectric layer. It may be noted that, so far, the optimization of the design parameters for a broadband absorber has been done only numerically, using full-wave simulations. Analytical modelling is restricted to calculating the Gap-plasmon dispersion to derive the width of the resonators required to have the resonance at a particular wavelength. In this section, we present a novel analytical framework to design a broadband MDM superabsorber for a general resonator-dielectric-substrate system by computing the required critical design parameters comprising of the resonator width (W) and height (H), and the dielectric thickness ($t_d$) as a function of operational wavelength ($\lambda$). This allows us to optimize the required design parameters for broadband absorption in a fully analytical approach.

**2.1. Gap plasmon dispersion**

Metallic nanoparticles have the unique ability of enhancing electromagnetic fields by confining light at sub-wavelength ranges through the excitation of surface plasmon polaritons (SPPs) at the metal-dielectric interface [42]. A Metal-Dielectric-Metal system, depicted in the insert of **Figure 1** can be modeled as a continuous layer gap plasmon resonators (CL-GPRs), where the electromagnetic field is localized within the dielectric gap due to magnetic coupling between the SPPs at the two metal-dielectric interfaces. The propagation constant $\beta$ of the Gap-plasmon satisfies the characteristic dispersion equation : [43]

$$\tanh\left(\frac{\sqrt{\beta^2 - k_0^2 \varepsilon_d}\, t_d}{2}\right) = \frac{-\varepsilon_d \sqrt{\beta^2 - k_0^2 \varepsilon_m}}{\varepsilon_m \sqrt{\beta^2 - k_0^2 \varepsilon_d}} \qquad (1)$$

Where $k_0$ is the wave vector of light in free space; $\varepsilon_m$ and $\varepsilon_d$ represent the dielectric constants for the metal and dielectric. **Equation 1** is solved numerically for a dielectric spacer thickness ($t_d$) of 50 nm (initial guess, which will be refined later). The gap-plasmon effective index ($n_{eff}$) is given by $\mathrm{Re}(\beta / k_0)$. Finally, the resonator width (W) is computed as a function of resonance wavelength (λ) using **Equation 2**, where *m* is the mode index, and $\phi$ is an additional phase shift.

$$W\left(\frac{2\pi}{\lambda}\right)n_{eff} = m\pi - \phi \qquad (2)$$

The Au ground plane is the significant contributor of absorption at lower wavelengths. The phase shift ($\phi$) upon reflection from metal is not a straightforward computation. Though for simplicity we assumed it to be zero, in practice, the non-zero phase shift results in a small (~20 nm) redshift of the resonance [44]. **Figure 1** depicts the dependency of the width of the resonators on the resonance wavelength for the fundamental mode ($m=1$). From **Figure 1**, 18 resonators with dimensions varying from 40 nm to 200 nm, in steps of 10nm, were selected to have a broadband resonance (see **Table 1a** for dimensions).

### 2.2. Supercell design

The gap plasmon dispersion gives only the characteristic width (W) of the resonators, and in principle is the same for any shape of the scattering object such as circles (diameter), squares (side) and crosses (limb). Although a relatively broadband response could be achieved by using only one type of resonator [25], a combination of different resonators assists in suppressing the interparticle coupling between two resonators which leads to dips in the absorptions spectrum. (see Supporting information). The proposed supercell design inspired by recent work[13] consists of *36* resonators, in a 6 x 6 grid, with 18 unique resonators consisting of squares (side: 40-90 nm), cylinders [25] (diameter: 100-150 nm), and crosses [13,45] (Limb; 140-200 nm, width 50 nm) (see **Figure 1**). The propagation length of SPP in *Au* is *~20 nm* at optical frequencies [46]. The center to center distance between two adjacent resonators is hence fixed at 250 nm thus guaranteeing a minimum distance between the edges of the elements that allows suppression of coupling between resonators. Polarization independence is achieved by placing the resonators in a *4-fold symmetry* as shown in **Table 1b**. The periodicity of the supercell is 1500 nm (6 x 250 nm).

## 2.3 Effective medium approximation

The supercell is then converted into an effective medium such that it can be modeled as a bulk medium of a certain effective permittivity [47]. This simplifies the MDM superabsorber greatly into a stacked layers structure, and the Transfer Matrix Method (described in the next sub-section) can be used to compute the effective resonator height and spacer thickness required for perfect absorption. The Asymmetric Bruggeman formulation is used to homogenize the supercell: the effective permittivity $\varepsilon_{eff}$ is computed at resonance using **Equation 3**. $\varepsilon_m$ and $\varepsilon_h$ are the permittivity of metal and host medium (air) respectively. The fill factor (f) for different resonator shapes is computed using **Equation 4**, where L = 1500 nm denotes the lattice period of the supercell, W is the characteristic width of the resonators and $W_1$ = 50 nm is the width of the crosses. It is to be noted here that for a given resonance wavelength, the contribution of non-resonating elements in the supercell are negligibly weak and hence they are ignored in the computation of effective permittivity. **Figure 2** depicts the computed effective permittivity. The real part of the permittivity is close to 1 (black curve), and the imaginary part of the permittivity (blue curve) is close to zero, confirming the free-space impedance matching.

$$\left(\frac{\varepsilon_m - \varepsilon_{eff}}{\varepsilon_m - \varepsilon_h}\right) = (1-f)\left(\frac{\varepsilon_{eff}}{\varepsilon_h}\right)^{\frac{1}{3}} \qquad (3)$$

$$f = \begin{cases} 2\left(\dfrac{W}{L}\right)^2 & squares \\ 2\pi\left(\dfrac{W}{2L}\right)^2 & circles \\ 2\left[\left(\dfrac{2WW_1}{L^2}\right) - \left(\dfrac{W_1}{L}\right)^2\right] & crosses \end{cases} \qquad (4)$$

## 2.4 Transfer Matrix Method

The MDM superabsorber can now be simplified as a stack of 2 layers of thickness $d_1 = H$ and $d_2 = t_d$ with permittivity $\varepsilon_{eff}$ and $\varepsilon_d$ respectively, embedded between an air superstrate and metal substrate with permittivity $\varepsilon_m$ (see inset **Figure 3(a)**). The relative permeability ($\mu_r$) of the entire system is considered to be 1. Inside the top homogenized layer and the bottom metal layer the waves are evanescent, and hence the wavevectors are purely imaginary. i.e. $k_1 = ik_0\sqrt{\varepsilon_{eff}}$ and $k_3 = ik_0\sqrt{-\varepsilon_m}$, whereas is the wavevector in the spacer layer. The Transfer Matrix (M) for such a system is defined as the product of scattering matrices for each layer (see [45] for derivations):

$$M = \begin{pmatrix} m_{11} & m_{12} \\ m_{21} & m_{22} \end{pmatrix} = S_0 S_1 S_2 S_3, \text{ where}$$

$$S_0 = \begin{pmatrix} 1 & 1 \\ k_0 & -k_0 \end{pmatrix}^{-1} \quad (air\ superstrate)$$

$$S_j = \begin{pmatrix} \cos(k_j d_j) & \frac{i}{k_j}\sin(k_j d_j) \\ ik_j \sin(k_j d_j) & \cos(k_j d_j) \end{pmatrix} \quad (layers\ \ j=1,2)$$

$$S_3 = \begin{pmatrix} 0 & e^{i(d_1+d_2)k_3} \\ 0 & -k_3 e^{i(d_1+d_2)k_3} \end{pmatrix} \quad (metal\ substrate)$$

The reflection and transmission amplitudes are given by:

$$r = \frac{\gamma_0 m_{11} + \gamma_0 \gamma_3 m_{12} - m_{21} - \gamma_3 m_{22}}{\gamma_0 m_{11} + \gamma_0 \gamma_3 m_{12} + m_{21} + \gamma_3 m_{22}} \text{ and } t = \frac{2\gamma_0}{\gamma_0 m_{11} + \gamma_0 \gamma_3 m_{12} + m_{21} + \gamma_3 m_{22}}.$$

Where $\gamma_j = \tilde{n}_j \cos\theta_j$, $\tilde{n}_j = n_j - ik_j$ is the refractive index of $j^{th}$ layer, and $\theta_j$ is the angle of incidence. Substituting the S- matrices into M and equating reflection and transmission amplitudes to zero gives **Equations 5, 6** for optimal spacer thickness ($t_d$) and resonator height ($H$) at normal incidence.

$$\cot\left(k_0\sqrt{\varepsilon_d}\,t_d\right) = \left(\frac{\sqrt{\frac{-\varepsilon_m}{\varepsilon_d}} + \sqrt{\frac{\varepsilon_d}{\varepsilon_{eff}}}}{\sqrt{\frac{-\varepsilon_m}{\varepsilon_{eff}}} - 1}\right) \quad (5)$$

$$\tan\left(ik_0\sqrt{\varepsilon_{eff}}\,H\right) = \sqrt{\frac{\varepsilon_{eff}}{-\varepsilon_m}} \left(\frac{\sqrt{\frac{-\varepsilon_m}{\varepsilon_d}}\tan\left(k_0\sqrt{\varepsilon_d}\,t_d\right) + 1}{\sqrt{\frac{\varepsilon_d}{-\varepsilon_m}}\tan\left(k_0\sqrt{\varepsilon_d}\,t_d\right) - 1}\right) \quad (6)$$

**Figure 3(a)** shows the computed values of the optimal spacer thickness ($t_d$) (black curve) and the optimal resonator height (H) (blue curve) as a function of resonance wavelength ($\lambda$). As seen in **Figure 3a**, the optimal spacer thickness ($t_d$) has a nearly linear dependence on the resonance wavelength. The ground *Au* plane is the significant contributor of absorption at shorter wavelengths. The optimal spacer thickness ($t_d$) of the supercell is derived as mean from 530-850 nm which gives a value of 42 nm, refining the initial guess of 50 nm . **Equation 6** is computed with this value for $t_d$. The optimal height of the resonators (blue curve) is greater at shorter wavelengths (as can be seen from **Figure. 3(a)**), but is relatively flat elsewhere. Again, taking into account the significant contribution of ground Au plane in absorption at shorter wavelengths, the height of the resonators is derived as 78 nm (mean from 600-850 nm). Hence, for broadband performance the optimal spacer thickness of the proposed MDM superabsorber is 42 nm, while the optimal height of the resonators in the supercell is 78 nm. Numerical simulations (elaborated in the next section) were performed to test the effectiveness of the proposed framework. The thickness of the spacer layer, and the height of the resonators were swept from 40 – 65 nm and 50 – 80 nm, respectively. **Figure 3b** shows the average absorption at normal incidence obtained in the 500 – 850 nm spectral range as a function of spacer thickness and resonator height. The absorption is maximum for a spacer thickness ($t_d$) of 45 nm, and for resonator height (H) of 75 nm, matching quite well with the analytical results.

## 3. Numerical simulation

The optical absorption of the proposed MDM superabsorber has been simulated with COMSOL Multiphysics® software. The structure consists of 36 *Au* resonators each of height 75 nm with dimensions given in **Table 1a**, arranged in a 6 x 6 supercell of periodicity 1500 nm as shown in **Table 1b**. The supercell is separated from the ground *Au* layer (thickness 200 nm) by a 45 nm *SiO₂* layer, and a 5 nm *Cr* adhesion layer. The edges of the resonators are rounded by 5 nm to account for imperfections in the fabrication. The Johnson and Christy permittivity model [48] is used for *Au* and *Cr*, and Malitson model [49] is used for $SiO_2$. In the 3D FEM simulation, Floquet boundary conditions are employed in both **x** and **y** directions, and perfectly matched layer boundary conditions are employed in the **z** direction. The 200 nm thick bottom *Au* layer is treated as an impedance matching layer (to reduce the computation time). The absorption is calculated as $A(\lambda) = 1 - R(\lambda)$, where $R(\lambda) = |S_{11}|^2$ is the reflection, as there was no transmission $T(\lambda) = |S_{21}|^2$ in the entire wavelength range. The simulated normal incidence absorption of the proposed MDM superabsorber structure for both TE (blue curve) and TM modes (red curve) fundamental modes in the 450-950 nm spectral range is shown in **Figure 4**. The insert depicts the schematic of the proposed MDM superabsorber. Although the analytical model was designed for the 500 – 850 nm spectral region, we can observe that the average absorption is above 0.97 for both polarizations in the 450 - 950 nm spectral region, with near unity absorption between 550 and 850 nm. The ground *Au* layer accounts for high absorption in the lower wavelength range, and the non-zero phase shift acquired upon reflection from the ground *Au* layer is attributed to the red shift. The dip around 800 nm, more prominent in the TE mode is attributed to a grating mode arising due to the periodicity. The simulated field patterns at various wavelengths are shown in **Supplementary Figure 2**.

The angular behaviour of the proposed MDM superabsorber is numerically simulated for both polarizations. **Figures 5a**, **5b** demonstrate that the MDM superabsorber has a weak polarization and launch angle dependence, and has high performance up to 45º (average absorption is 0.896 for TM and 0.943 for TE). The dips in the spectrum are attributed to the grating orders of the supercell, which become more prominent at higher angles. The angle dependent absorption of the superabsorber is depicted in **Figure 5c, 5d** for both polarizations from 12º to 24º (step size 0.3º). The dispersion is almost flat, and the dip centred around 850 nm is attributed to a grating mode of the supercell.

## 4. Experimental Section

The proposed MDM superabsorber structure was fabricated on a silicon substrate using electron beam lithography. The active area of the fabricated device is 300 μm × 300 μm. A Silicon substrate is chosen to provide the necessary mechanical support. A flexible polymer substrate can also be used [50]. A 200 nm thick gold film, along with a 5 nm Chromium adhesion layer was first deposited on the silicon wafer using DC Magnetron sputtering, followed by chemical vapor deposition of a 45 nm thick $SiO_2$ film. A 200 nm thick PMMA 459 A4 resist layer was then spin coated. The resist was exposed by an electron beam with 20 kV accelerating voltage and a 10 μm focusing aperture. The exposed resist was developed in a 1:3 solution of MIBK: IPA for 90s and rinsed in IPA for 60s. A 5 nm Chromium adhesion layer, and a 75 nm gold layer was then sputtered sequentially on the patterned resist. The resist was later removed by acetone lift-off. A false colored scanning electron microscopy (SEM) image of the sample is shown in Figure 6 (inset).

The polarization and angle dependent reflection of the fabricated sample shown in **Figure 6** is measured using a collimated (<0.3° divergence) halogen lamp (450 – 950 nm) source with 120 μm spot and a USB-2000 fiber spectrometer (Ocean Optics). The angle of incidence is limited to 20°– 45° due to mechanical constraints of the setup. Sample reflection spectra were source normalized, from which we derived the sample extinction. As there was no measurable transmittance and non-specular reflection in the entire spectral range studied, it is assumed that the extinction is equal to the absorbance. At 20° incident angle, we report an average absorption of over 98% for both polarizations in the 450-950 nm spectral region. Average absorption measured for various incidence angles match well with simulations, and is tabulated in **Table 2**. The uncertainty in the reported experimental data is 0.05%. The artifacts/inhomogeneity in the device fabrication explains the marginally better experimental performance, and the broadening of the dips in the spectrum (more pronounced at 45°). Device fabrication can further be improved by implementing strategies reported in [51]. Numerical simulations were carried out to model the effects of inhomogeneity in crosses: all the crosses in the supercell were replaced by diamonds of appropriate dimensions (See Supporting Information – Figure 1). Although the experimentally realized crosses have significant inhomogeneity, the experimental response matches better with simulations having crosses rather than diamonds.

## 5. Conclusions

A novel analytical framework for the design of a broadband, polarization independent superabsorber is presented in this work. The analytical design results are substantiated by numerical simulation results. We report an average polarization independent absorbance of ~97% in the 450 – 950 nm bandwidth spanning the entire optical region. Experimentally, an absorbance of over 98% in the 450-950 nm spectral region at 20° launch angle has been demonstrated. These results are in a good agreement with simulations. The presented design framework can be applied for any wavelength region. It can be extended for higher angles and higher modes. Although intended for normal incidence and fundamental modes, this framework is the first step in having an all analytical approach to design and optimize MDM superabsorbers.

**Acknowledgements**

The authors would like to thank Dr. Man Xu, Dr. Sachin Kasture, and Dr. Benjamin Brenny for discussions and comments on the manuscript. This work is funded by TU/e through the MELISSA project.

**Figure 1.** Width (W) of resonator as a function of resonance wavelength (λ) computed by numerically solving the Gap plasmon dispersion equation for the Metal-Dielectric-Metal system shown in the inset.

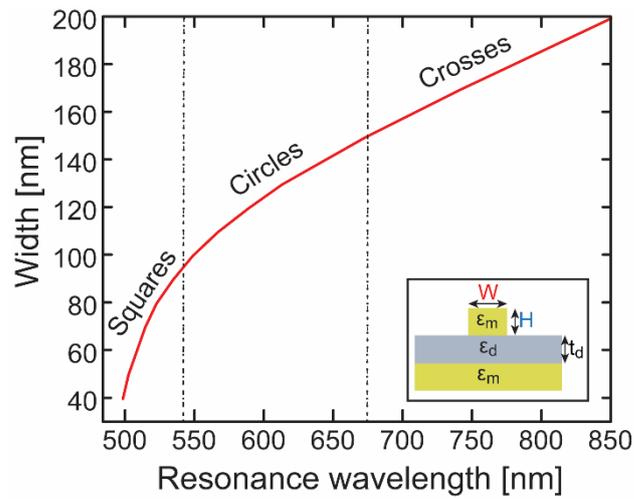

**Figure 2.** Homogenizing the permittivity of the supercell by Asymmetric Bruggeman formulation. The effective permittivity is computed at resonance for each of the resonators. The real part of the effective permittivity is close to 1 (black curve), while the imaginary part is negligibly small (blue curve).

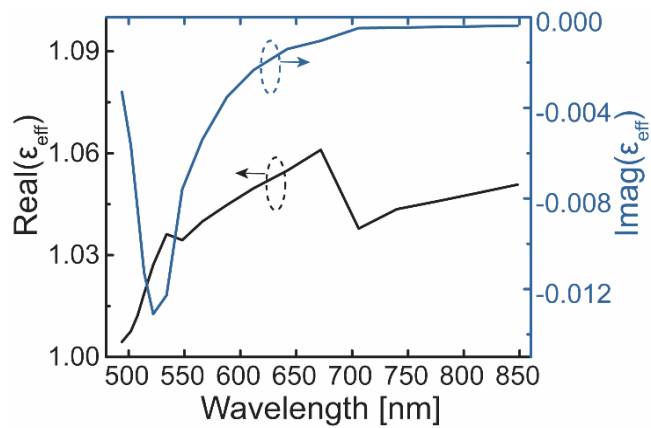

**Figure 3.** Optimizing the Spacer thickness and the Resonator height of the supercell.
**a)** Transfer-Matrix method results to get the optimal Spacer thickness ($t_d$) (black curve) and the Resonator height (**H**) (blue curve) required for perfect absorption at different wavelengths in a 3-layer stack shown in the insert. **b)** Average absorption of the proposed MDM superabsorber structure in the 500-850 nm spectral range for varying spacer thickness ($t_d$) and Resonator height (**H**) obtained using full-wave simulations. The arrow indicates the optimal value of 45 nm and 75 nm for spacer thickness and resonator height, respectively.

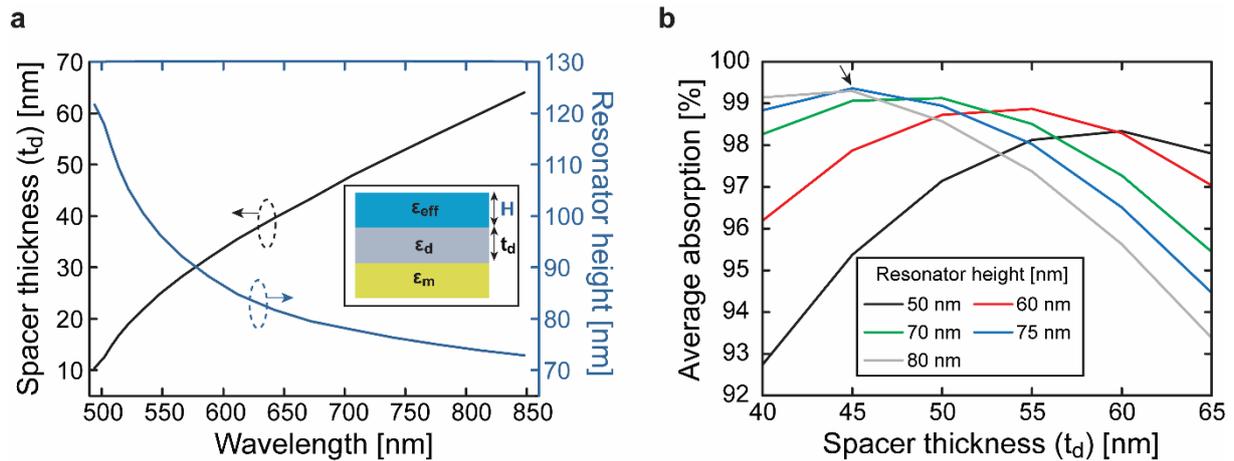

**Figure 4.** Simulated normal incidence absorption for the TE Mode (Blue curve) and TM Mode (Red curve) for the proposed MDM superabsorber structure shown in the inset.

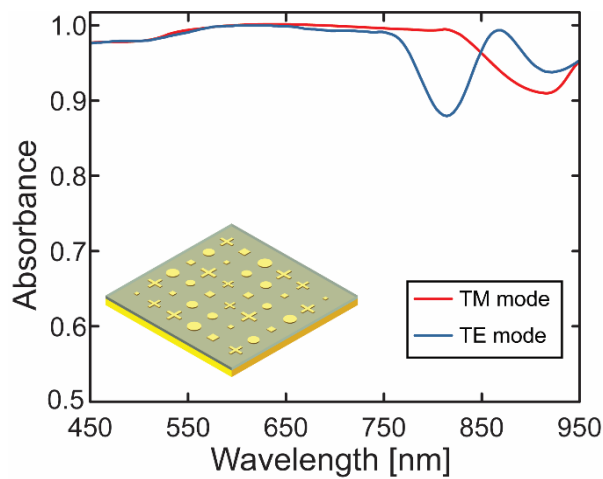

**Figure 5.** Simulated angle dependent absorption of the proposed MDM superabsorber structure for the TM Mode **(a)** and TE Mode **(b)**. Dispersion for the TM and TE Modes are shown in **c)** and **d)** respectively. The dip in the spectrum centered around 850 nm is the grating order, which becomes more pronounced at higher launch angles.

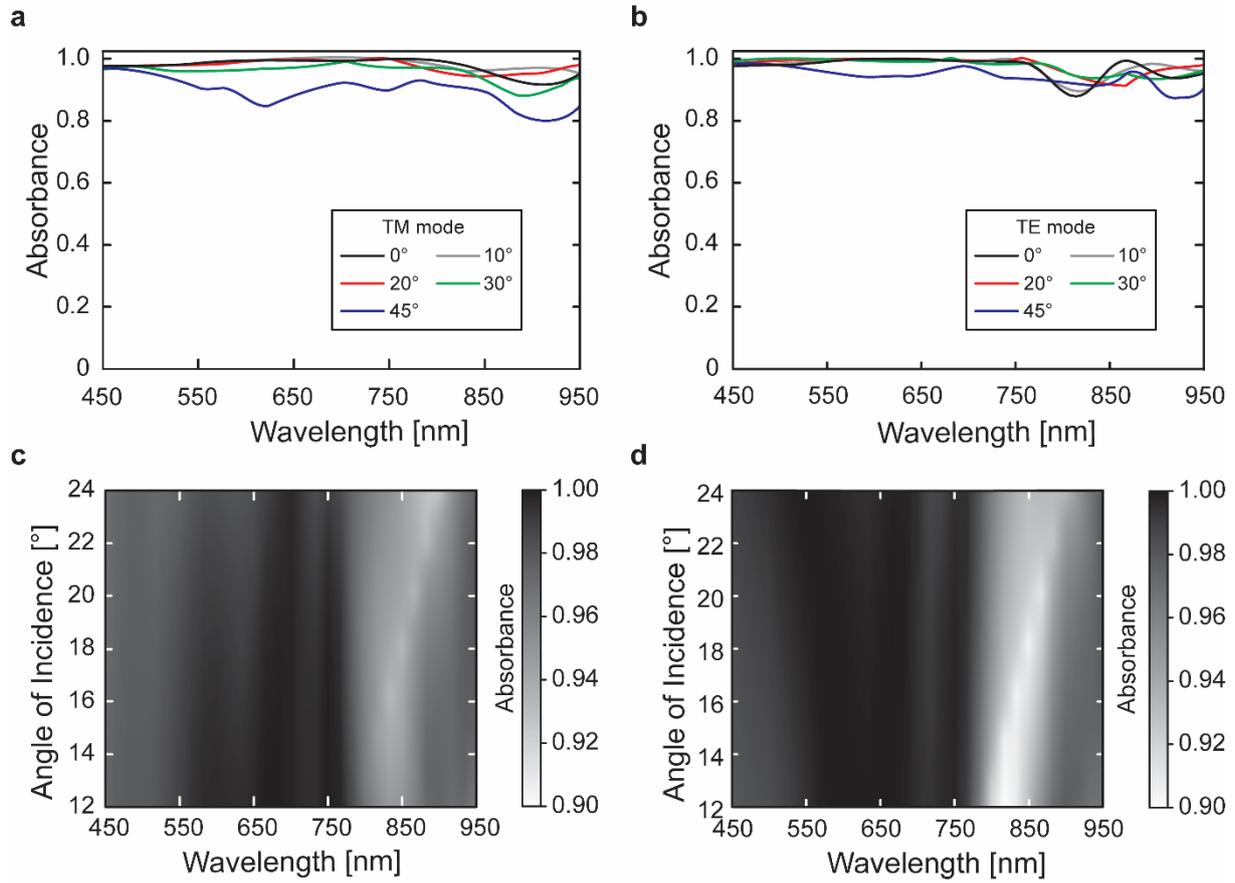

**Figure 6.** The measured angle dependent absorption $A = 1 - R$ of the fabricated device for the TM Mode **(a)** and TE mode **(b)**. The insert shows the false colored SEM image of the fabricated device. Squares, Circles and Crosses are coded by Yellow, Red, and Blue, respectively. The white scale bar represents 200 nm.

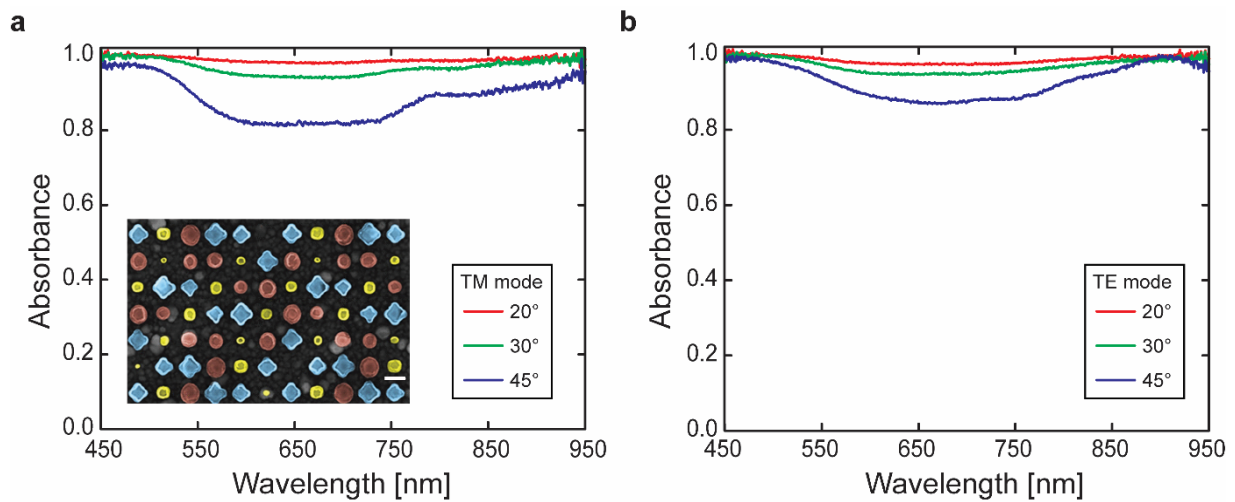

**Table 1a.** Dimensions of individual resonators used in the design.

| Number | Square length (nm) (SQ) | Cylinder diameter (nm) (C) | Cross Limb (nm) (CR) |
|---|---|---|---|
| 1 | 40 | 100 | 140 |
| 2 | 50 | 110 | 160 |
| 3 | 60 | 120 | 170 |
| 4 | 70 | 125 | 180 |
| 5 | 80 | 140 | 190 |
| 6 | 90 | 150 | 200 |

**Table 1b**. Placement of the resonators in the supercell.

| SQ1 | CR4 | C5  | SQ4 | C4  | CR3 |
|---|---|---|---|---|---|
| CR2 | SQ3 | C1  | CR6 | SQ2 | SQ6 |
| CR5 | C3  | SQ5 | CR1 | C2  | C6  |
| C6  | C2  | CR1 | SQ5 | C3  | CR5 |
| SQ6 | SQ2 | CR6 | C1  | SQ3 | CR2 |
| CR3 | C4  | SQ4 | C5  | CR4 | SQ1 |

**Table 2.** Average absorption in the 450 – 950 nm spectral range.

| Angle of Incidence [°] | Numerical Simulation | | Experiment | |
|---|---|---|---|---|
| | **TM Mode** | **TE Mode** | **TM Mode** | **TE Mode** |
| 0  | 97.92 % | 97.30 % | *       | *       |
| 20 | 97.64 % | 97.89 % | 98.85 % | 98.60 % |
| 30 | 95.64 % | 97.67 % | 96.73 % | 96.97 % |
| 45 | 89.60 % | 94.30 % | 88.30 % | 92.92 % |

# Supporting Information

**Title** A Broadband Superabsorber at Optical Frequencies: Design and Demonstration

Arvind Nagarajan[*], Kumar Vivek, Manav Shah, Venu Gopal Achanta, and Giampiero Gerini

**Figure S1:** Comparison of absorbance of the proposed supercell (black line) with one consisting of cylinders of appropriate dimensions (see inset) (red line), and one where crosses are replaced with diamond shaped elements (see inset) (blue line), at normal incidence. Clearly, resonators with different shapes help in suppressing the reflection.

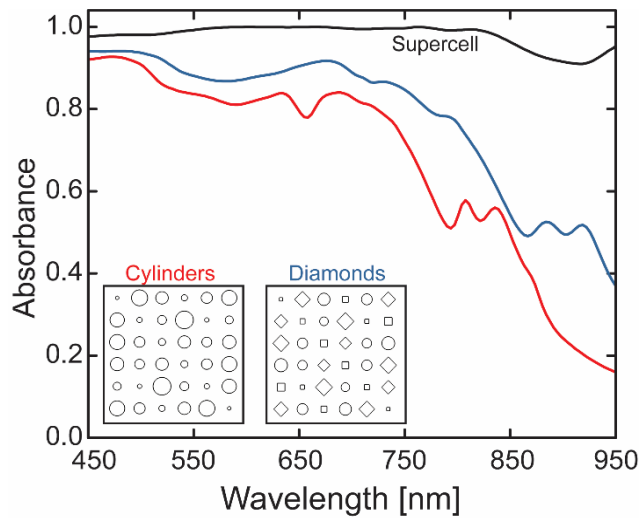

At shorter wavelengths, all resonators contribute to the overall absorption, with the smallest square resonators providing maximum contribution as shown in Figure S2a for $\lambda = 500$ nm. Medium-sized resonators i.e. cylinders are excited and dominate the absorption process at intermediate wavelengths, as shown in Figure S2b for $\lambda = 700$ nm. At $\lambda = 900$ nm only the larger cross resonators are excited – the distributions of the electric fields are shown in Figure. S2c. It can be observed that the energy is confined within these structures for the wavelength specified. The point to be noted here is, at wavelengths ($\leq 500$nm) gold substrate is penetrated more by the light field due to small resonator dimensions and the overall absorption performance is high due to inter-band components of gold permittivity.

**Figure S2:** Simulated moduli of electric field inside the dielectric of the broadband MDM superabsorber for impinging wavelengths of 500 nm (a), 700 nm (b), and 900 nm (c). Units of electric field is V/m.

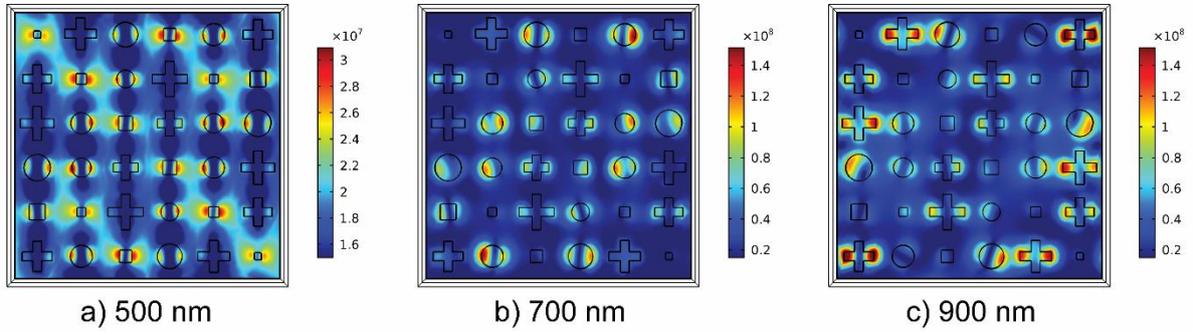